\documentclass[twocolumn,showpacs,preprintnumbers,amsmath,amssymb]{revtex4}
\usepackage{graphicx}% Include figure files
\usepackage{dcolumn}% Align table columns on decimal point
\usepackage{rotating}%
\usepackage{amssymb}%
\usepackage{amsmath}
\usepackage{color}
\usepackage{bm}% bold math
\usepackage{epsfig}

\begin{document}

\preprint{APS/123-QED}

\title{Decomposing the dynamics of heterogeneous delayed networks \\
with applications to connected vehicle systems}

\author{R\'obert Szalai$^1$ and G\'abor Orosz$^2$}%
\affiliation{$^1$Department of Engineering Mathematics, University of Bristol, Bristol, BS8 1TR, UK \\
$^2$Department of Mechanical Engineering, University of Michigan, Ann Arbor, Michigan 48109 USA} %%%

%\author{R\'obert Szalai}%
%\email{r.szalai@bristol.ac.uk}%
%\affiliation{Department of Engineering Mathematics, University of Bristol,
%Bristol, BS8 1TR, UK}

%\author{G\'abor Orosz}
%\email{orosz@umich.edu} %%%
%\affiliation{Department of Mechanical Engineering, University of Michigan,
%Ann Arbor, Michigan 48109 USA}

\date{\today}

\begin{abstract}
Delay-coupled networks are investigated with nonidentical delay times and the
effects of such heterogeneity on the emergent dynamics of complex systems are
characterized. A simple decomposition method is presented that decouples the
dynamics of the network into node-size modal equations in the vicinity of
equilibria. The resulting independent components contain distributed delays
that map the spatiotemporal complexity of the system to the time domain. We
demonstrate that this new approach can be used to reveal new physical
phenomena in heterogenous vehicular traffic when vehicles are linked via
vehicle-to-vehicle (V2V) communication.
\end{abstract}

\pacs{02.30.Ks, 05.45.Xt, 64.60.aq, 89.75.Kd, 89.40.Bb}% PACS, the Physics and Astronomy
\maketitle

%%%%%%%%%%%%%%%%%%%%%%%%%%%%%%%%%%%%%%%%%%%%%%%%%%%%%%%%%%%%%%%%%%%
%\section{Motivation}
%%%%%%%%%%%%%%%%%%%%%%%%%%%%%%%%%%%%%%%%%%%%%%%%%%%%%%%%%%%%%%%%%%%
%\vspace{-5mm}

%{\bf Motivation}
The dynamics of delayed networks are in the current interest of research
communities in physics, biology and engineering. Applications include neural
networks \cite{CamKob12,KanYan13,Oro13_1}, gene regulatory networks
\cite{JosLopOttShiBen11,OroMoeMur10}, semiconductors lasers
\cite{SorGarMirFis13,FluYanDahSch10}, and traffic systems
\cite{OroWilSzaSte09,OroWilSte10,TreKesHel06}. In these systems, delays arise
in the couplings between components due to finite-time information
propagation, which greatly influence the arising patterns of activity.
However, in the above cases it has been assumed that the delays are
identical. This is clearly not the case in physical systems, where
communication channels have different transmission rates and information
travels greatly varying distances. In this Rapid Communication, we
characterize the behavior of realistic heterogenous delayed systems about
equilibria by applying a simple decomposition method. In particular, we
analyze the dynamics of a connected vehicle system and show that having an
optimal level of delay heterogeneity may maximize stability of the uniform
flow which has significant implications on traffic dynamics.

In order to understand the system-level behavior arising through delayed
connectivity, large systems of delay differential equations have to be
analyzed. Even in the absence of delays, one needs to handle high-dimensional
systems. Moreover, delays make the dynamics infinite dimensional which
typically leads to complicated dynamics even for simple systems. For the case
of identical delays, decomposition methods have been proposed
\cite{OlfMur04,FluYanDahSch10,Hod10,HunKorSzy10,CepOlg11,Oro13_1} to
investigate the dynamics in the vicinity of the synchronized equilibrium,
which result in (linear autonomous) delayed modal equations of small size.
The decomposition methods have been extended to handle the dynamics in the
vicinity of synchronous periodic orbits using Floquet theory \cite{Oro13_1}
and synchronous chaos using Lyapunov exponents \cite{KinEngReeZigKan09}.
Further developments allow decomposition of the dynamics in the vicinity of
steady and oscillatory cluster states \cite{ChoDahHovSch10,Oro13_2} and the
analysis of traveling wave solutions \cite{KanYan13}. However, networks with
heterogenous delays escaped many attempts of modal decomposition, because no
finite dimensional transformation can untangle the interaction of
nonidentical delays. As a first step, in this Rapid Communication we propose
a simple approach that can handle heterogenous delays in the vicinity of
equilibria at the linear level. The key idea is to decompose the system in
the Laplace domain and then transform the uncoupled modal equations back to
the time domain. This results in delayed modal equations with distributed
delays where the spatiotemporal complexity of the original coupled system is
mapped to the time domain by the delay distributions.

As a motivating example we consider a simple, but heterogeneous car-following
model \cite{Hel01,TreKesHel06,OroWilSte10} where the interaction of vehicles
is facilitated by automatic control that is based on wireless
vehicle-to-vehicle (V2V) communication \cite{Cav10,SafPil12}. As different
channels of information exchange have naturally different delay times, this
model is unsuitable for available modal decomposition techniques. By applying
our method, we illustrate how individual delays are mixed in the decomposed
system. The corresponding modes are traveling wave-like solutions that become
traveling waves of different wave lengths in the case of identical delays and
next-neighbor interactions \cite{OroWilSzaSte09,OroMoeBul10}. Synthesizing
the results obtained for individual traffic modes, we develop a systematic
understanding of heterogeneous traffic dynamics for our simple example. The
presented method may be used to develop control strategies for larger systems
involving V2V communication.

%%%%%%%%%%%%%%%%%%%%%%%%%%%%%%%%%%%%%%%%%%%%%%%%%%%%%%%%%%%%%%%%%%%
%\section{Decomposing infinite dimensional systems into finite number of modes}
%%%%%%%%%%%%%%%%%%%%%%%%%%%%%%%%%%%%%%%%%%%%%%%%%%%%%%%%%%%%%%%%%%%
%\vspace{-5mm}

%{\bf Decomposition method}
We consider a general description of dynamics on a network written in the form of
\begin{equation}\label{eq:nonlinmodel}
\dot{x}_i(t) = f\big(x_i(t)\big) +  \sum_{j=1}^N a_{ij}\, g\big(x_i(t),x_j(t-\tau_{ij})\big)\,, %%%
\end{equation}
for $i=1,\ldots, N$, where the state of node $i$ is given by the vector $x_i
\in \mathbb{R}^n$, the internal dynamics are described by $f(x_i)$, and the
couplings $g(x_i,x_j)$ depend on the states of the interacting nodes
\cite{note1}. The time delays $\tau_{ij}$ account for signal propagation and
processing times. The coupling structure of the system is captured by a
weighted directed graph described by the $N$-dimensional adjacency matrix
$A_N = [a_{ij}]$ whose elements are defined as $a_{ij} \neq 0$ if node $j$ is
connected to node $i$ and $a_{ij} = 0$ otherwise for $i,j = 1,\ldots,N$. Our
goal is to decompose the dynamics of (\ref{eq:nonlinmodel}) around an
equilibrium and define modal coordinates in which the system becomes
uncoupled and the corresponding modal equations can be analyzed separately by
current state-of-the-art tools \cite{InsSte11,RooSza07,Ata03}.

In this paper, for the sake of simplicity, we focus on the dynamics in the
vicinity of the synchronous/uniform equilibrium $x_i(t) \equiv x^*$,
$i=1,\ldots,N$ \cite{note2}. We define the perturbations $y_i = x_i - x^*$
for $i=1,\ldots,N$, so the linearization of (\ref{eq:nonlinmodel}) can be
written as
\begin{equation}\label{eq:linmodel}
\dot{y}_i(t) = L\, y_i(t) + R \sum_{j=1}^N a_{ij}\, y_j(t-\tau_{ij})\,. %%%
\end{equation}
The $n$-dimensional matrices $L,R$ are given by
\begin{equation}\label{eq:LR}
L = \partial f\big(x^*\big) + m\, \partial_1 g\big(x^*,x^*\big)\,, %%%
\quad
R =  \, \partial_2 g\big(x^*, x^*\big)\,, %%%
\end{equation}
where $\partial_1$ and $\partial_2$ represent partial derivatives with
respect to the first and second set of variables, respectively, while $m =
\sum_{j=1}^N a_{ij}$ is the (constant) row sum \cite{note2}.

Using the notation ${\bf y} = {\text{col}}\,[\,y_1\ \ y_2\ \ \ldots\ \ y_N\,]
\in \mathbb{R}^{nN}$ the linear system (\ref{eq:linmodel}) can be rewritten
as
\begin{equation}\label{eq:linmodelformal}
\dot{\bf y}(t) = (I_N \otimes L)\, {\bf y}(t) + (\mathcal{A}_N \otimes R)\, {\bf y}(t)\,, %%%
\end{equation}
where $I_N$ is the $N$-dimensional identity matrix while $\mathcal{A}_N =
[a_{ij}\,\mathcal{S}_{-\tau_{ij}}]$ is an adjacency operator that
incorporates the components of the adjacency matrix as well as the time-shift
operator
\begin{equation}\label{eq:shiftoperator}
\mathcal{S}_{-\tau_{ij}} y_j(t) = y_j(t-\tau_{ij})\,.
\end{equation}

In order to decompose system (\ref{eq:linmodelformal}) into $N$ modes of size
$n$, one needs to diagonalize the adjacency operator $\mathcal{A}_N$.
First, we take the Laplace transform of (\ref{eq:linmodelformal}) and neglect the terms that would arise from a particular initial condition:
\begin{equation}\label{eq:linmodelformallaplace}
s\,{\bf Y}(s) = \big(I_N \otimes L\big)\, {\bf Y}(s) + \big(B_N(s) \otimes R\big)\, {\bf Y}(s) \,, %%%
\end{equation}
where the matrix $B_N(s) = [a_{ij}\, {\rm e}^{-s\,\tau_{ij}}]$  is the
Laplace transform of the adjacency operator $\mathcal{A}_N$
\cite{YiNelUls10,note3}. Then we define the modal transformation
\begin{equation}\label{eq:trafo}
{\bf Y}(s) = \big(T_N(s) \otimes I\big){\bf Z}(s)\,,
\end{equation}
where the columns of the matrix $T_N(s)$ consist of the eigenvectors of
$B_N(s)$. This yields
\begin{equation}\label{eq:modaleqformal}
s\,{\bf Z}(s) = \big(I_N \otimes L\big)\, {\bf Z}(s) + \big(C_N(s) \otimes R\big)\, {\bf Z}(s) \,, %%%
\end{equation}
where ${\bf Z}(s)$ is the Laplace transform of the vector ${\bf z} =
{\text{col}}\,[\,z_1\ \ z_2\ \ \ldots\ \ z_N\,] \in \mathbb{R}^{nN}$ and the
diagonal matrix $C_N(s)$ contains the eigenvalues $\Lambda_k(s)$ of $B_N(s)$.
That is, the node-size modal equations in the Laplace domain become
uncoupled:
\begin{equation}\label{eq:modaleq}
s\,Z_k(s) =  L\, Z_k(s) + R\, \Lambda_k(s)\, Z_k(s) \,, %%%
\end{equation}
for $k=1,\ldots, N$. We remark that even if the adjacency matric $A_N$ is not
diagonalizable (i.e., it has eigenvalues whose algebraic multiplicity is
larger than their geometric multiplicity), the matrix $B_N(s)$ in
(\ref{eq:linmodelformallaplace}) may still be diagonalized, that is,
heterogeneity in the delays can destroy the symmetry imposed by the coupling
structure.

The inverse Laplace transform of (\ref{eq:modaleq}) results in the
distributed delay systems
\begin{equation}\label{eq:modaleqtime}
\dot z_k(t) = L\, z_k(t) + R \int_0^t \lambda_k(\xi)\, z_k(t-\xi) {\rm d} \xi \,, %%%
\end{equation}
in the time domain where $\lambda_k(\xi)$ is the inverse Laplace transform of
$\Lambda_k(s)$ for $k=1,\ldots, N$. We remark that the infinite
dimensionality of transformation $T_N(s)$ in (\ref{eq:trafo}) can be
understood by observing that it ``shuffles" the present and past values of
the coordinates in the time domain.

The stability of the modal equations (\ref{eq:modaleqtime}) can be analyzed
by the direct methods given in \cite{Ste89} or by obtaining $\lambda_k(\xi)$
using inverse Laplace transform. Note that the eigenvalues can be written as
\begin{equation}\label{eq:lambdas}
\Lambda_k(s) = \bar{\Lambda}_k \big( {\rm e}^{-s\, \tau_{11}}, {\rm e}^{-s\, \tau_{12}},\ldots, {\rm e}^{-s\, \tau_{NN}} \big)\,. %%%
\end{equation}
In general $\bar{\Lambda}_k$ is a nonlinear function and ${\rm e}^{-s\,
\tau_{ij}}$ are periodic along the contour $s = {\rm i} \omega$, which makes
(\ref{eq:lambdas}) quasi-periodic with frequencies $\tau_{ij}$. Furthermore,
the inverse Laplace transform is equivalent to the Fourier transform
\cite{Kuh13}, that takes quasi-periodic functions into sums of periodic
functions with frequencies from the set $\Omega = \{\sum_{ij} p_{ij}\,
\tau_{ij} \ge 0 : p_{ij} \in \mathbb{Z}\}$. Therefore the eigenvalues can be
approximated as
\begin{equation}\label{eq:modaleigs}
\Lambda_k(s) \approx \sum_{\ell \colon T_{k,\ell}\in\Omega} \rho_{k,\ell}\, {\rm e}^{-s\, T_{k,\ell}}\,,
\end{equation}
where the coefficients are calculated by truncating the Fourier transform
\begin{equation}\label{eq:modalweights}
\rho_{k,\ell} = \lim_{T\to\infty} \frac{1}{T} \int_0^T \Lambda_k({\rm i} \omega)\, {\rm e}^{{\rm i} \omega\, T_{k,\ell}} {\rm d} \omega\,. %%%
\end{equation}

If $\bar{\Lambda}_k$ in (\ref{eq:lambdas}) is a smooth function of its
variables, more insight can be gained by using multi-variable Taylor
expansion about a point where all the variables assume the value ${\rm
e}^{-s\, T_0}$ (identical delays $\tau_{ij} = T_0$) \cite{Sch05}. The
corresponding coefficients can be obtained by calculating the partial
derivatives $\partial^{q_1}_1 \cdots \partial^{q_M}_M \bar{\Lambda}_k \big(
{\rm e}^{-s\, T_0}, \ldots, {\rm e}^{-s\, T_0} \big) = {\rm e}^{(Q-1)s\, T_0}
\Phi_{k,q_1\cdots q_M}$ where $M=N^2$ is the number of variables, $Q = q_1 +
\cdots + q_M$ is the order of the derivative, and $\Phi_{k,q_1\ldots q_M}$
only depend on the coupling strengths $a_{ij}$, hence they are independent of
$s$ \cite{supp}. Choosing $T_0 = \min\{\tau_{ij}\}$ guarantees that all the
resulting exponential terms are in the form of ${\rm e}^{-s\, T_{k,\ell}}$
with non-negative $T_{k,\ell} \in \Omega$, which is required by causality.
The clear advantage of the Taylor expansion over the integral method is that
it is more likely to provide analytical results in some simple cases.

The inverse Laplace transform of ${\rm e}^{-s\, T_{k,\ell}}$ is the Dirac
delta $\delta(\xi- T_{k,\ell})$, that is, (\ref{eq:modaleigs}) results in the
distribution
\begin{equation}\label{eq:modaleigstime}
\lambda_k(\xi) \approx \sum_{\ell \colon T_{k,\ell}\in\Omega} \rho_{k,\ell}\, \delta (\xi - T_{k,\ell})\,.
\end{equation}
Thus, the convolution integral in (\ref{eq:modaleqtime}) can be evaluated
yielding the delay equations
\begin{equation}\label{eq:modaleqtimetaylor}
\dot z_k(t) = L\, z_k(t) + R \sum_{\ell \colon T_{k,\ell}\in\Omega}  \rho_{k,\ell}\, z_k(t-T_{k,\ell}) \,, %%%
\end{equation}
for $k=1,\ldots,N$, which approximate distributed delays by (infinitely many)
discrete delays. The stability of the equilibrium can be studied using the
approximate characteristic equations
\begin{equation}\label{eq:modalchareqs}
{\rm det}\Big( s I - L - R \sum_{\ell \colon T_{k,\ell}\in\Omega}  \rho_{k,\ell}\, {\rm e}^{-s\,T_{k,\ell}}\Big)=0 \,, %%%
\end{equation}
where $I$ is the $n$-dimensional identity matrix. Considering $s={\rm
i}\,\omega$, $\omega \geq 0$ the stability boundaries can be derived
analytically \cite{InsSte11}, while discretizing time in
(\ref{eq:modaleqtime}) or (\ref{eq:modaleqtimetaylor}) may allow numerical
approximation of the characteristic roots $s$ \cite{RooSza07}.

%%%%%%%%%%%%%%%%%%%%%%%%%%%%%%%%%%%%%%%%%%%%%%%%%%%%%%%%%%%%%%%%%%%
%\section{Connected vehicle dynamics}
%%%%%%%%%%%%%%%%%%%%%%%%%%%%%%%%%%%%%%%%%%%%%%%%%%%%%%%%%%%%%%%%%%%
%\vspace{-5mm}

%{\bf Connected vehicle dynamics}
To gain insight to the physics of connected vehicle systems and to illustrate
our above derived formalism we consider a simple car-following model.
Car-following models describe the motion of individual vehicles moving in
continuous time and space \cite{Hel01,OroWilSte10}. These models can be
extended to incorporate vehicle-to-vehicle (V2V) communication which changes
the network structure by introducing long range connections with heterogenous
delays.

\begin{figure}[t!]
\begin{center}

\epsfig{file=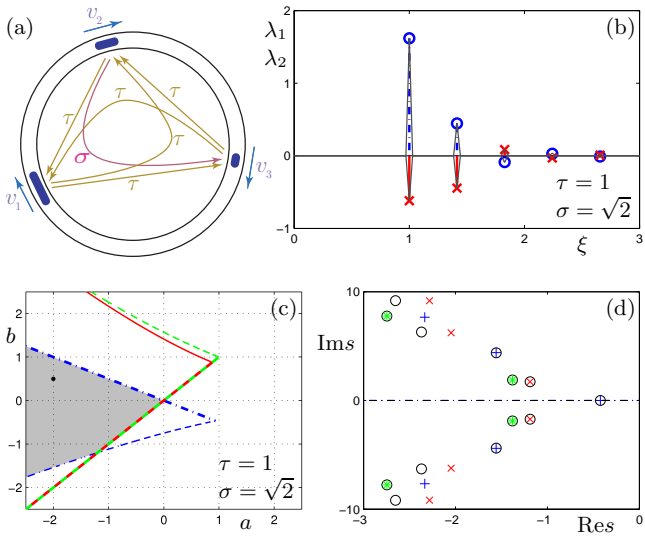,width=86.0mm}

\vspace{-2mm}

\end{center}
\caption{\label{fig:fig01} (Color online) (a) A sketch of three vehicles
following each other on a ring-road. Orange and purple arrows show the
direction of information propagation through V2V communication with delays
marked on each link. (b) Delay distributions (\ref{eq:modaleigstime}) with
support (\ref{eq:carfolsupport}) for modes $k=1,2$. Dash-dotted blue lines
with circles and solid red lines with crosses correspond to the distributions
calculated analytically for modes 1 and 2, respectively. Numerical
approximations are shown as solid gray curves. (c) Stability chart with the
stable domain shaded. Dash-dotted blue, solid red and dashed green curves
correspond to the 1st, 2nd and 3rd modes, respectively.  When crossing thin
curves stability changes through a pair of complex conjugate characteristic
roots while thick lines correspond to stability change with zero
characteristic root. (d) Comparing the leading characteristic roots obtained
from (\ref{eq:carfollin}) (black circles) and (\ref{eq:carfoldecomposed})
(blue plus, red cross, green star for modes 1,2,3) for $K=6$. Parameters
correspond to the point marked by a dot on panel (c).}

\vspace{-5mm}

\end{figure}

We consider a simplified model with node dimension $n=1$, where only the
vehicles' speed $v_i$, $i=1,\ldots N$ are exchanged via communication:
\begin{equation}\label{eq:carfolnonlin}
\dot v_i(t) = \gamma\big(v_0 - v_i(t)\big) + \sum_{j=1}^N \beta_{ij} V\big(v_j(t-\tau_{ij}) -v_i(t) \big)\,.
\end{equation}
Here $V$ is a monotonously increasing function with $V(0)=0$, while $\gamma$
and $\beta_{ij}$ represent the gains to maintain the desired velocity $v_0$
and zero relative velocity, respectively. We consider periodic boundary
conditions, i.e., put the vehicles on a ring road. It can be shown
analytically that the linear stability conditions obtained for large $N$ are
equivalent to the conditions that guarantee attenuation of perturbations
along vehicle platoons \cite{OroWilSte10}. For the sake of simplicity here we
restrict ourselves to the simplest nontrivial case of $N=3$ vehicles; see
Fig.~\ref{fig:fig01}(a). Also, we consider the coupling constants
$\beta_{ii}=0$ and $\beta_{ij} = \beta$ for $i \neq j$ and the heterogenous
delay setup $\tau_{ij} = \tau$ for $i \neq j$ except $\tau_{32} = \sigma \geq
\tau$. This mimics the scenario that vehicle 1 obstructs the transmission of
information from vehicle 2 to vehicle 3, resulting in longer delay.

By linearizing (\ref{eq:carfolnonlin}) about the uniform equilibrium $v_i(t)
\equiv v_0$, $i=1,2,3$ we obtain the linear system
\begin{equation}\label{eq:carfollin}
\begin{bmatrix}
\dot{\tilde v}_1(t) \\ \dot{\tilde v}_2(t) \\ \dot{\tilde v}_3(t)
\end{bmatrix}
=
a
\underbrace{\begin{bmatrix}
1 & 0 & 0 \\
0 & 1 & 0 \\
0 & 0 & 1
\end{bmatrix}}_{I_N}
\begin{bmatrix}
\tilde{v}_1(t) \\ \tilde{v}_2(t) \\ \tilde{v}_3(t)
\end{bmatrix}
+ b
\underbrace{\begin{bmatrix}
0 & \mathcal{S}_{-\tau} & \mathcal{S}_{-\tau} \\
\mathcal{S}_{-\tau} & 0 & \mathcal{S}_{-\tau} \\
\mathcal{S}_{-\tau} & \mathcal{S}_{-\sigma} & 0
\end{bmatrix}}_{\mathcal{A}_N}
\begin{bmatrix}
\tilde{v}_1(t) \\ \tilde{v}_2(t) \\ \tilde{v}_3(t)
\end{bmatrix}
\end{equation}
where $\tilde{v}_i = v_i - v_0$, $a = -\gamma - 2\beta V'(0)$ and $b = \beta
V'(0)$.

To decompose (\ref{eq:carfollin}) into its modal components we calculate the
Laplace transform of the adjacency operator $\mathcal{A}_N$:
\begin{equation}\label{eq:BNS}
B_N(s) =
\begin{bmatrix}
0 & {\rm e}^{-s\,\tau} & {\rm e}^{-s\,\tau} \\
{\rm e}^{-s\,\tau} & 0 & {\rm e}^{-s\,\tau} \\
{\rm e}^{-s\,\tau} & {\rm e}^{-s\,\sigma} & 0
\end{bmatrix}\,,
\end{equation}
cf.~(\ref{eq:linmodelformallaplace}), which possesses the eigenvalues
\begin{equation}\label{eq:carfolmodaleigs}
\begin{split}
&\Lambda_{1,2}(s) = \textstyle \frac{1}{2} \big( {\rm e}^{-s\,\tau} %%%
\pm \sqrt{5{\rm e}^{-2s\,\tau} + 4 {\rm e}^{-s\,(\tau+\sigma)}} \big)\,,
\\
&\Lambda_3(s) = -{\rm e}^{-s\,\tau}\,,
\end{split}
\end{equation}
that appear in the modal equations (\ref{eq:modaleq}) in the Laplace domain.
To approximate the convolution by discrete delays in the modal equations
(\ref{eq:modaleqtime}) in the time domain we calculate the Taylor expansion
of $\bar{\Lambda}_{1,2}(x,y) = \frac{1}{2} \big( x \pm \sqrt{5 x^2 + 4 x y}
\big)$ about $(x_0,y_0)$ and then set $x_0=y_0 = {\rm e}^{-s\,\tau}$;
cf.~(\ref{eq:lambdas}) and the discussion after (\ref{eq:modalweights}). The
resulting expression is a polynomial in ${\rm e}^{-s\,\tau}$ and ${\rm
e}^{-s\,\sigma}$. Calculating the inverse Laplace transform we obtain the
modal equations
\begin{eqnarray}\label{eq:carfoldecomposed}
&&\dot{\tilde w}_k(t) = a\, \tilde{w}_k(t) + b \sum_{\ell=0}^K \rho_{k,\ell}\, \tilde{w}_k(t-T_{k,\ell})\,, \qquad k=1,2\,, \nonumber %%%
\\
&&\dot{\tilde w}_3(t) = a\, \tilde{w}_3(t) - b\, \tilde{w}_3(t-\tau)\,,
\end{eqnarray}
where the support of the delay distributions is given by
\begin{equation}\label{eq:carfolsupport}
T_{k,\ell} = \ell \sigma - (\ell-1)\tau\,,
\end{equation}
for $k=1,2$. In Fig.~\ref{fig:fig01}(b) dash-dotted blue lines with circles
and solid red lines with crosses show the delay distributions for modes 1 and
2, respectively, for $K=6$. These analytical results are approximated very
well by the distributions obtained numerically using (\ref{eq:modalweights})
that are shown as solid gray curves. We used $T = 1000\,\tau$ and $p_1,p_2 =
-20,\ldots,20$ so that $T_{k,\ell} = p_1 \tau + p_2 \sigma \geq 0$.

The modal equations (\ref{eq:carfoldecomposed}) result in the characteristic
equations
\begin{equation}\label{eq:carfolchareqs}
\begin{split}
&s - a - b \sum_{\ell=0}^K \rho_{k,\ell}\, {\rm e}^{-s\,T_\ell} = 0\,, \qquad k=1,2\,, %%%
\\
&s - a + b\, {\rm e}^{-s\,\tau} = 0\,.
\end{split}
\end{equation}
Substituting $s={\rm i}\,\omega$, $\omega \geq 0$ we obtain the stability
boundaries in the $(a,b)$-plane in parametric form for each mode. These are
shown in Fig.~\ref{fig:fig01}(c) where modes are distinguished by color and
line type (see caption). Thin curves correspond to $\omega>0$ while thick
lines correspond to $\omega=0$. When crossing a thin curve, oscillations
arise with frequency $\omega$ while crossing a thick line leads to
non-oscillatory stability loss. (In the corresponding nonlinear system
(\ref{eq:carfolnonlin}), Hopf and fold bifurcations take place.) Indeed, the
equilibrium is stable if all modes are stable as indicated by the shaded
domain. The accuracy of the stability boundaries improve when increasing the
number of delays $K$ in (\ref{eq:carfolchareqs}).

%%%%%%%%%%%%%%%%%%%%%%%%%%%%%%%%%%%%%%%%%%%%%%%%%%%%%%%%%%%%%
%%%%%%%%%%%%%%%%%%%%%%%%%%%%%%%%%%%%%%%%%%%%%%%%%%%%%%%%%%%%%

When discretizing time in systems (\ref{eq:carfollin}) and
(\ref{eq:carfoldecomposed}), one may calculate the characteristic roots $s$
numerically. Fig.~\ref{fig:fig01}(d) compares the characteristic roots for
$K=6$ for the parameter values corresponding to the dot in panel (c). Circles
represent the characteristic roots obtained for (\ref{eq:carfollin}) while
other symbols represent the characteristic roots obtained for the individual
modes in (\ref{eq:carfoldecomposed}); see caption. Notice that the leading
characteristic roots are reproduced very well while deviations occur for
characteristic roots with smaller negative real part \cite{note4}.

%%%%%%%%%%%%%%%%%%%%%%%%%%%%%%%%%%%%%%%%%%%%%%%%%%%%%%%%%%%%%
\begin{figure}[t!]
\begin{center}

\epsfig{file=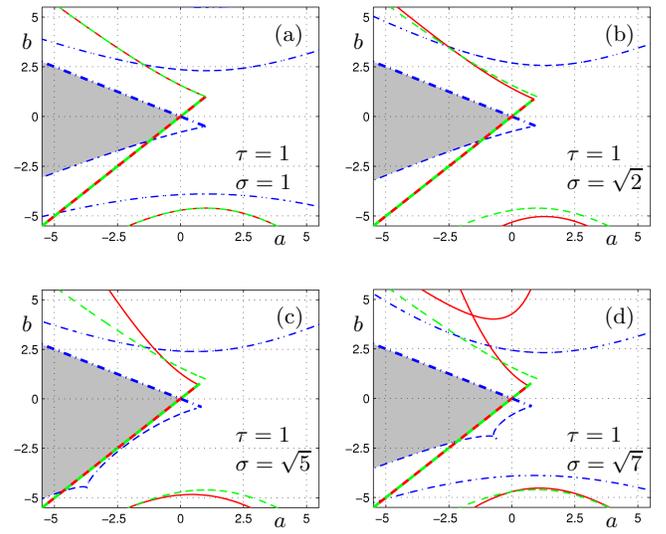,width=86.0mm}

\vspace{-2mm}

\end{center}
\caption{\label{fig:fig02} (Color online) Stability charts with different
delay values as indicated on each panel. The same notation is used as in
Fig.~\ref{fig:fig01}(c).}

\vspace{-5mm}

\end{figure}
%%%%%%%%%%%%%%%%%%%%%%%%%%%%%%%%%%%%%%%%%%%%%%%%%%%%%%%%%%%%%

In order to evaluate the effects of delay heterogeneity on the system
dynamics we depict the stability charts for different $\sigma$ values in
Fig.~\ref{fig:fig02}. Using a larger parameter window compared to
Fig.~\ref{fig:fig01}(c) reveals other stability curves (belonging to higher
values of $\omega$). Crossing these only makes the system ``more unstable"
with more characteristic roots on the right hand side. As can be seen in
Fig.~\ref{fig:fig02}(a) the stability boundaries for modes 2 and 3 become
identical when $\sigma=\tau$. In this case mode 1 is called tangential mode:
when instability occurs in this mode the synchronous/uniform configuration is
kept. On the other hand, modes 2 and 3 are transversal modes: stability
losses in these modes breaks synchrony \cite{Oro13_1}, leading to traveling
waves. Such categorization is not possible for heterogenous delays. In this
case, each mode gives a different set of curves (see
Fig.~\ref{fig:fig02}(b,c,d)) that correspond to different spatiotemporal
patterns: traveling waves that are asymmetric due to the delays. Notice that
as the heterogeneity in the delays increases the stable domain may increase
or decrease in the $(a,b)$ parameter plane. In fact, choosing the level of
delay heterogeneity appropriately one may maximize the stable domain and so
increase the robustness of the uniform flow.

%%%%%%%%%%%%%%%%%%%%%%%%%%%%%%%%%%%%%%%%%%%%%%%%%%%%%%%%%%%%%%%%%%%
%\section{Discussion}
%%%%%%%%%%%%%%%%%%%%%%%%%%%%%%%%%%%%%%%%%%%%%%%%%%%%%%%%%%%%%%%%%%%
%\vspace{-5mm}

%{\bf Discussion}
In summary, this Rapid Communication introduced a new method in analyzing
complex systems with coupling-delay heterogeneity that can be followed in a
range of applications. In the vicinity of the synchronized equilibrium,
through an infinite-dimensional modal transformation, modal equations with
distributed delays were derived for heterogeneous delayed networks so that the
spatiotemporal complexity of the network is embedded in the delay
distributions. The analysis of the modal equations provide a systematic way
to map out the system-level dynamics. It was demonstrated that the method can
be used to analyze the spatiotemporal dynamics of connected vehicle systems.
It was found that having appropriate level of delay heterogeneity can
maximize the robustness of the uniform traffic flow.

Our future research we will extend these results using more realistic
car-following models and connectivity structures. Indeed, applications extend
beyond this specific problem. For example, one may extend the current
framework to non-synchronized equilibria that can be used to design gene
regulatory circuits of given functional properties \cite{OroMoeMur10}. Also,
extending the framework to periodic orbits may allow one to characterize
self-organized criticality in neural networks \cite{Oro13_1,Oro13_2}, which
can lead to better understanding of neuro-computation and memory in the
brain.

\quad
\\
\quad
%%%%%%%%%%%%%%%%%%%%%%%%%%%%%%%%%%%%%%%%%%%%%%%%%%%%%%%%%%%%%%%%%%%
%\begin{acknowledgments}
The authors greatly acknowledge the suggestions of Tam\'as Insperger, G\'abor
St\'ep\'an, and the two anonymous referees. We also thank Wubing Qin for his
help with one of the figures.
%\end{acknowledgments}
%%%%%%%%%%%%%%%%%%%%%%%%%%%%%%%%%%%%%%%%%%%%%%%%%%%%%%%%%%%%%%%%%%%

%%%%%%%%%%%%%%%%%%%%%%%%%%%%%%%%%%%%%%%%%%%%%%%%%%%%%%%%%%%%%%%%%%%
\bibliographystyle{apsrmp4-1}        % this should be delated before submission
\bibliography{biblio_traffic}   % Produces the bibliography via BibTeX.

\begin{thebibliography}{33}
\bibitem [{not(N1)}]{note1}%
  \BibitemOpen
   (\bibinfo {year} {N1}),\ \href@noop {} {}\bibinfo {note} {The decomposition
  method is presented for the case where the node dynamics $f$ are and the
  couplings $g$ are uniform to reduce the algebraic complexity. Generalization
  for heterogeneous $f$ and $g$ will follow in future work.}\BibitemShut
  {Stop}%
\bibitem [{not(N2)}]{note2}%
  \BibitemOpen
   (\bibinfo {year} {N2}),\ \href@noop {} {}\bibinfo {note} {The synchronized
  equlibrium $x^*$ can be obtained by solving the algebraic equation $0 =
  f\big(x^*\big) + m\, g\big(x^*,x^*\big)$. The row sum $m = \sum_{j=1}^N
  a_{ij}$ must be the same for every $i$ to ensure the existence of synchronous
  solutions.}\BibitemShut {Stop}%
\bibitem [{not(N4)}]{note4}%
  \BibitemOpen
   (\bibinfo {year} {N4}),\ \href@noop {} {}\bibinfo {note} {When the number of
  delays $K$ is increased the leading characteristic roots are reproduced with
  higher accuracy, but accuracy may decrease for characteristic roots with
  smaller real part.}\BibitemShut {Stop}%
\end{thebibliography}
%%%%%%%%%%%%%%%%%%%%%%%%%%%%%%%%%%%%%%%%%%%%%%%%%%%%%%%%%%%%%%%%%%%

\end{document}